# HSH-carbon: A novel $sp^2$-$sp^3$ carbon allotrope with an ultrawide energy gap

Jia-Qi Liu[1], Qian Gao[1], Zhen-Peng Hu[1,†]

[1]*School of Physics, Nankai University, Tianjin 300071, China*

†*Corresponding author. Email: zphu@nankai.edu.cn*

**Abstract:**

A $sp^2$-$sp^3$ hybrid carbon allotrope named HSH-carbon is proposed by the first-principles calculations. The structure of HSH-carbon can be regarded as a template polymerization of [1.1.1]propellane molecules in a hexagonal lattice, as well as, an AA stacking of recently reported HSH-$C_{10}$ consisting of carbon trigonal bipyramids. Based on calculations, the stability of this structure is demonstrated in terms of the cohesive energy, phonon dispersion, Born-Huang stability criteria, and ab initio molecular dynamics. HSH-carbon is predicted to be a semiconductor with an indirect energy gap of 3.56 eV at the PBE level or 4.80 eV at the HSE06 level. It is larger than the gap of Si and close to the gap of c-diamond, which indicates HSH-carbon is potentially an ultrawide bandgap semiconductor. The effective masses of carriers in the VB and CB edge are comparable with wide bandgap semiconductors such as GaN and ZnO. The elastic behavior of HSH-carbon such as bulk modulus, Young's modulus and shear modulus is comparable with that of T-carbon and much smaller than that of c-diamond, which suggests that HSH-carbon would be much easier to be processed than c-diamond in practice.

**Keywords**

First-principles calculation, novel carbon allotropes, pentagonal ring

## 1. Introduction

As one of the most fundamental elements for life on the earth, carbon could generate allotropes and compounds of great diversity with various and unique properties owing to its strong ability to bond with other elements in sp-, $sp^2$- and $sp^3$-hybridized forms. The three ways of hybridization construct various carbon structures, such as sp for one dimensional carbon chains (carbyne) [1, 2], $sp^2$ for graphite [3] and

sp$^3$ for diamond. It is expected to construct structures with the combination and arrangement of these different ways of hybridization, which could lead to a variety of novel carbon allotropes with exotic and attractive properties.

A series of studies on novel carbon allotropes with special characteristics have been carried out and made splendid achievements in the past few decades. Apart from graphene [4], fullerenes [5] and carbon nanotubes [6] as representative achievements, a considerable number of novel carbon allotropes have been designed, studied, predicted and synthesized over the past decades, such as sp$^3$-hybridized superhard C-Centered orthorhombic allotrope C$_8$ (Cco-C$_8$) [7], graphyne and graphdiyne of sp and sp$^2$ atoms with high degrees of π-conjunction, uniformly distributed pores and tunable electronic properties [8, 9], and an expanded cubane with a C$_{56}$ core of sp-sp$^3$ hybridization [10], etc. As a quite successful representative of sp$^3$-hybridized novel carbon allotropes, T-carbon was revealed in 2011, which was obtained by substituting a carbon tetrahedron as the building block for each atom in diamond [11-13]. Subsequently derivative structures from T-carbon were also proposed by introducing different building blocks [14]. For sp$^2$-sp$^3$ hybridization, some novel periodic structures were reported, such as penta-graphene [15, 16] and pentadiamond [17, 18]. It is noticed that those structures are composed of pentagonal rings as a remarkable building motif. Inspired by T-carbon, maybe people could find a three-dimensional motif rather than planar pentagons as a novel building block for sp$^2$-sp$^3$ carbon allotropes.

Recently, a new two-dimensional carbon allotrope HSH-C$_{10}$ was proposed by our group [19], which can be considered as a mixture of honeycomb and star lattices. From a different point of view, HSH-C$_{10}$ can also be obtained by substituting each atom in graphene with a trigonal bipyramid, where the trigonal bipyramid is just a three-dimensional motif rather than planar pentagons. Simply, with the AA stacking of HSH-C$_{10}$, a structure named HSH-carbon is proposed as a novel three-dimensional carbon allotrope consisting of sp$^2$ and sp$^3$ hybridized carbon atoms, which can also be considered as a polymerization of [1.1.1]propellane molecules. Based on the first-principles calculations, the stability, electronic and mechanical properties of HSH-carbon are analyzed in this article. HSH-carbon is predicted to be a potential ultrawide bandgap semiconductor, which would be easier to be processed than c-diamond.

## 2. Methods:

All calculations are performed on the basis of density functional theory as implemented in the Vienna ab initio simulation (VASP) package [20]. The projector augmented-wave (PAW) method is adopted to describe the interactions between electrons and ions [21]. Generalized gradient approximation (GGA) with the Perdew-Burke-Ernzerhof (PBE) functional is used for the exchange-correlation interaction between electrons [22]. The plane-wave energy cutoff is set to 520 eV. Sampling over Brillouin zone is carried out with a Γ-centered 7×7×13 grid. The geometric optimization is accomplished when the Hellmann-Feynman force on each ion is less than 0.001 eV/Å. A self-consistent calculation of electrons converges when the energy difference between adjacent electronic steps is less than $1\times10^{-5}$ eV, and for phonon dispersion calculation the convergence criterion is $1\times10^{-8}$ eV. The open source package Phonopy is used for phonon dispersion calculations [23]. Ab initio NpT ensemble simulations are performed in a Langevin thermostat on a p(3×3×3) super cell (270 atoms) for 10 ps at 500K and 800K with a timestep of 2 fs and only Γ sampling in K-space.
The cohesive energy per atom is defined as:

$$E_{coh} = (E_{bulk} - nE_{atom})/n \qquad (1)$$

where $E_{bulk}$ is the energy of the condensed structure, $E_{atom}$ is the energy of the isolated atoms and $n$ is the number of atoms in the cell.

## 3. Results and Discussions:

3.1 Structure

Specific structure information of HSH-carbon is summarized in Table 1. Based on the symmetry of P6/mmm space group, the carbon atoms can be divided into two types, which are represented by the different Wyckoff Positions. They are marked as C1 and C3, respectively. There are three different bond lengths between them, where the values of C1-C1, C1-C3 and C3-C3 bonds are 1.484, 1.547 and 1.318 Å, respectively. The bond length of C3-C3 (1.318 Å) is a little shorter than 1.337 Å of the C-C double bond in ethylene but much longer than 1.207 Å of the C-C triple bond in ethyne, which reflects a double bond nature of the C3-C3 bond and the $sp^2$ hybridization of C3 atoms. The bond lengths of C1-C3 and C1-C1 are respectively longer than 1.502 Å of the intra-tetrahedron bond and 1.417 Å of the inter-tetrahedron bond in T-carbon, indicating a single bond nature and the $sp^3$ hybridization of C1 atoms. In a word, HSH-carbon is a $sp^2$-$sp^3$ carbon allotrope.

The top and side views of HSH-carbon are presented in Fig.1 (a) and (b), respectively. As shown in Fig. 1(a), C1 atoms at the corner and C3 atoms on the edge together constitute the honeycomb lattice. As shown in Fig. 1(b), the bond angle $\theta_{131}$ is 78.9°, which is much smaller than the $sp^2$ bond angle of 120° in ethylene and 133.4° in $sp^2$-$sp^3$ hybrid pentadiamond [17]. The $\theta_{113}$ of 129.4° and $\theta_{313}$ of 84.0° are also different from the $sp^3$ bond angle of 109.5° in methane. All these data suggest that the hybridization here is not as typical as the general $sp^2$ and $sp^3$ hybridization, which could be a result of the three-dimensional trigonal bipyramid as a building motif.

As shown in Fig. 1, there are many channels with different diameters in HSH-carbon, especially the channels with a diameter about 7.42 Å along c axis (Fig. 1(a)), which could reduce the density of this material. The calculated density of HSH-carbon is 1.62 g cm$^{-3}$, which is close to that of T-carbon (1.50 g cm$^{-3}$) and much smaller than that of c-diamond (3.52 g cm$^{-3}$). The porous structure and low density will enable HSH-carbon to have potential applications in molecular separation, hydrogen storage [11], and lithium storage [14].

3.2 Stability

The function of total energy per atom versus volume per atom for HSH-carbon, T-carbon, and c-diamond is shown in Fig. 2(a). From the plot, it can be observed that each curve has a single minimum, which indicates they all have a stable equilibrium. And the lower total energy suggest that HSH-carbon is stabler than T-carbon. The cohesive energy of HSH-carbon is calculated to be -7.00 eV/atom, which is lower than T-carbon (-6.66 eV/atom) but higher than c-diamond (-7.83 eV/atom). This also suggests HSH-carbon is stabler than T-carbon, though it is not as stable as c-diamond. Since T-carbon has been successfully synthesized [24], the synthesis of HSH-carbon in experiments can be expected. In principle, this would not be a hard work, as the practical synthesis can be achieved by a template polymerization using [1.1.1]propellane as the precursor.

The phonon dispersion has been calculated to evaluate the dynamical stability of HSH-carbon. As shown in Fig. 2(b), there is no imaginary frequency in the phonon spectra along the high symmetry directions, suggesting a good dynamical stability of the material. The highest frequency of 55.483 THz (1851 cm$^{-1}$) is higher than that of T-carbon (1760 cm$^{-1}$) and c-diamond (1303 cm$^{-1}$) [11], as HSH-carbon has the shortest bond (C3-C3) among those three materials. There is a direct band gap of 8.578 THz (286 cm$^{-1}$) at Γ and an indirect band gap of 6.708 THz (224 cm$^{-1}$) between Γ and A in Fig. 2(b), which results in double gaps in DOS of phonon. And these features could be useful for identifying the material in experiments.

Alternatively, five independent elastic constants $C_{11}, C_{33}, C_{44}, C_{12}, C_{13}$ are calculated to be 251. 51 GPa, 532.90 GPa, 55.89 GPa, 214.22 GPa and 22.32 GPa for HSH-carbon, respectively. And these values satisfy the Born-Huang elastic stability criteria for a mechanically stable hexagonal system [25, 26]:

$$C_{11} > 0, C_{44} > 0, C_{11} + C_{22} - 2C_{13}^2 / C_{33} > 0, C_{66} = (C_{11} - C_{12})/2 > 0 \qquad (2)$$

This also suggests that HSH-carbon is a mechanically stable carbon allotrope.

Furthermore, ab initio molecular dynamics (AIMD) simulations using NpT ensemble in a Langevin thermostat have also been performed to examine the thermal stability of HSH-carbon. To reduce the restriction of the periodic boundary condition and probe into the possible structure reconstruction, a p(3×3×3) super cell containing 270 atoms is built for the simulations. The simulations are performed for 10 ps with a timestep of 2 fs at 500K and 800K, and the average potential energy fluctuation and structures after the simulations are as shown in Fig. 2(c) and (d), respectively. Clearly, the average potential energy under the temperature of 500K and 800K does not fluctuate greatly and there is no structure reconstruction in both cases, though the cell parameters may vary slightly. These results suggest that HSH-carbon is thermally stable for a temperature as high as 800K.

## 3.3 Electronic properties

The electronic band structure of HSH-carbon is shown in Fig. 3(a). It can be observed that the conduction band minimum (CBM) is located at K and the valence band maximum (VBM) is at A, resulting in an indirect band gap 3.56 eV at the PBE level. It is larger than the gap of HSH-$C_{10}$ (2.89 eV) [19] as the dangling bonds are saturated by the AA stacking. The band gap can be even larger in practice, as normal DFT results usually underestimate the gap. And a value of 4.80 eV can be achieved when the HSE06 functional is applied. With its band gap larger than 3.4 eV of GaN and close to the gap of c-diamond, HSH-carbon has the potential to be an ultrawide bandgap semiconductor. Generally, the breakdown field increases as the band gap increases for semiconductors [27], so HSH-carbon may possess higher breakdown fields than GaN, enabling it to tolerate larger voltages for potential high-power electronic applications.

As shown in Table 2, the anisotropic effective masses of electrons in the conduction band edge and holes in the valence band edge are obtained from the band structure. The effective masses of electrons at the CBM are a little larger than that of c-diamond ($0.57m_0$) [28] and those near the $\Gamma$ point are comparable with those of GaN ($0.33m_0$ for $\Gamma\rightarrow M$, $0.36m_0$ for $\Gamma\rightarrow K$) [29]. The effective masses of holes in the VB edge are comparable with those of GaN (-1.58 $m_0$ for $\Gamma\rightarrow K$, -1.93 $m_0$ for $\Gamma\rightarrow M$, -2.03 $m_0$ for $\Gamma\rightarrow A$) and smaller than wide-bandgap ZnO (-4.31 $m_0$ for $\Gamma\rightarrow K$, -4.94 $m_0$ for $\Gamma\rightarrow M$, -1.98 $m_0$ for $\Gamma\rightarrow A$) [29]. To this point, with comparable effective masses of carriers and a band gap larger than GaN and ZnO, HSH-carbon can be considered as a good candidate of the third-generation semiconductors.

As shown in Fig. 3(b), $p_x$ ($p_y$) component dominates the band edge states around the band gap in the projected density of states (PDOS) of HSH-carbon, suggesting there would be a π binding system. The local density of states (LDOS) shown in the inset confirms this π binding nature at the band edges. LDOS of the highest branch of valence bands (VB) at $\Gamma$ point shows that π bonds are dominating, while π* bonds are dominating at the lowest branch of conduction bands (CB) at $\Gamma$ point. And this further confirms the $sp^2$ hybridization of C3 atoms, as $p_z$ participates in the $sp^2$ hybridization to form σ or σ* bond and the out-$sp^2$-plane component ($p_x$ or $p_y$ depending on the coordination) forms π or π* bond. And this conjugated π-bonding system explains the relatively small effective masses of holes near Γ point.

Alternatively, electron localization function (ELF) is also implemented to identify the bonding type of HSH-carbon. The ELF of HSH-carbon is compared with a partial methylated [1.1.1]propellane molecule. As shown in Fig. 3(c) and (d), both ELF are almost identical. As the bonding type in the partial methylated [1.1.1]propellane is clear, it can be concluded the bonds between C1-C1 and C1-C3 atoms are single bonds while the C3-C3 bonds are double bonds in HSH-carbon. Again, it confirms the $sp^2$ hybridization of C3 atoms and $sp^3$ hybridization of C1 atoms.

3.4 Mechanical properties

The ultrahigh mechanical parameters make c-diamond difficult and expensive to be used in devices in practice, which slows down its application in manufacturing. As discussed above HSH-carbon is potentially an ultrawide bandgap semiconductor, mechanical properties are studied to see whether it is easy to be processed. With the Voigt-Reuss-Hill averaging scheme [30, 31], the bulk modulus, Young's modulus, shear modulus and Poisson's ratio of HSH-carbon turn out to be 170 GPa, 120 GPa, 43 GPa and 0.384, respectively (Table 3). As shown in Table 3, except the Poisson's ratio, the mechanical parameters of HSH-carbon are comparable to those of T-carbon, but much smaller than those of c-diamond. This illustrates that the stiffness under load and the rigidity to the shear stress of HSH-carbon are much lower than c-diamond. It thus can be expected that HSH-carbon would be much easier to be processed than c-diamond in practice.

As HSH-carbon is anisotropic, the anisotropic elastic properties are also calculated with Elastic Anisotropy Measures (ElAM1.0) [32]. The projection along the c axis of 3D graphical representation of the elastic anisotropy of HSH-carbon are shown in terms of Young's modulus (Fig. 4(a)), shear modulus (Fig. 4(b)), linear compressibility (Fig. 4(c)) and Poisson's ratio (Fig. 4(d)), respectively. As elastic properties are rotationally symmetric with respect to the c axis, the projection graphs only show a cross section along the c axis. From Fig. 4, the maximum and the minimum of those elastic parameters can be easily observed, which are summarized in Table 4. As shown in Table 4, the Young's modulus has a maximum of 530.76 GPa along the c axis and a minimum of 69.03 GPa in its perpendicular plane, showing a remarkable anisotropy with the $A_E$ ratio of 7.69. It implies that the stiffness is especially higher along the c axis. The shear modulus, the linear compressibility, and the Poisson's ratio also show the clear anisotropy, which indicates one can easily process HSH-carbon along a specific direction.

## 4. Conclusion:

For a summary, we propose a novel $sp^2$-$sp^3$ hybridized carbon allotrope building up by the motif of trigonal bipyramid base on the first-principles calculations, which is named HSH-carbon. HSH-carbon contains two types of atoms arranging with the P6/mmm space group. HSH-carbon is confirmed to be energetically, dynamically, mechanically, and thermally stable by means of cohesive energy, phonon dispersion, Born-Huang criteria, and AIMD simulations, respectively. It has a low density of 1.62 g cm$^{-3}$ with a porous structure, suggesting potential applications in molecular separation, hydrogen storage, and lithium storage. It has an indirect band gap of 3.56 eV from the PBE results (4.80eV from the HSE06 results), and the effective masses of carriers in the VB and CB edge are comparable with wide bandgap semiconductors such as GaN and ZnO, indicting HSH-carbon would be an ultrawide bandgap semiconductor. The elastic properties such as bulk modulus, Young's modulus and shear modulus are much smaller than those of c-diamond, suggesting that HSH-carbon would be much easier to be processed than c-diamond in practice. This work provides a new point of using three-dimensional motif for the design of novel $sp^2$-$sp^3$ carbon allotropes, and experimental verifications are welcome.

**Acknowledgements**

This work was supported by the National Natural Science Foundation of China (No. 12134019, 21773124), the Fundamental Research Funds for the Central Universities Nankai University (No. 63221346, 63213042), the Supercomputing Center of Nankai University (NKSC), and the Prop plan from Hongzhiwei Technology.

Tables and Figures

**Table 1.** Structure data of HSH-carbon.

| Structure Information | | |
|---|---|---|
| Space group | | P6/mmm |
| a(Å) | | 6.422 |
| b(Å) | | 6.422 |
| c(Å) | | 3.450 |
| α(deg) | | 90.000 |
| β(deg) | | 90.000 |
| γ(deg) | | 120.000 |
| Wyckoff Position | C1(4h) | (0.333, 0.667, 0.215) |
| | C3(6k) | (0.441, 0.559, 0.500) |
| $d_{C-C}$(Å) | C1-C1, C1-C3, C3-C3 | 1.484, 1.547, 1.318 |

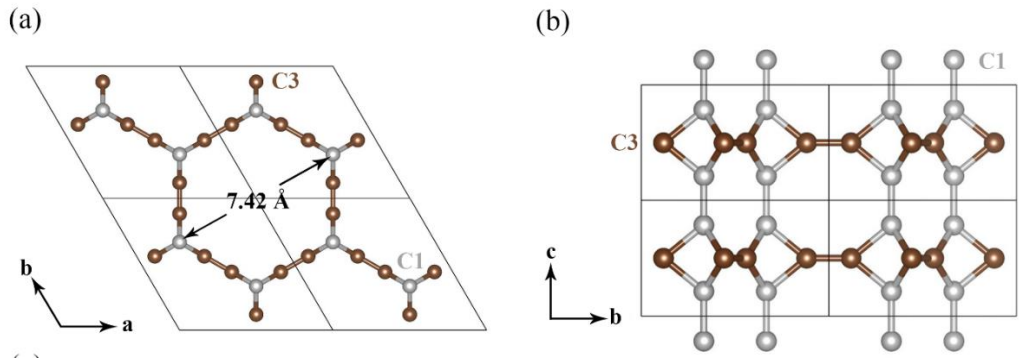

**Fig. 1 (a)** Top view and **(b)** side view of HSH-carbon. The solid lines represent the periodical boundary.

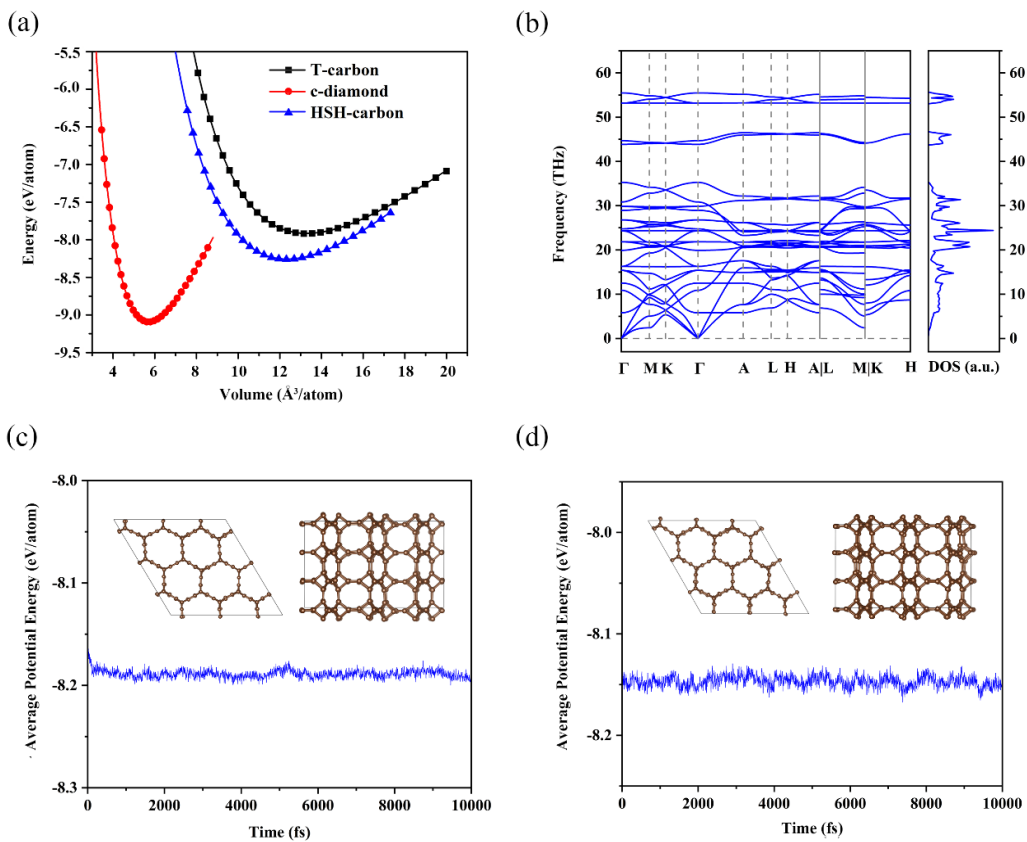

**Fig. 2 (a)** The total energy per atom as a function of volume per atom function for T-carbon, c-diamond and HSH-carbon, respectively. **(b)** The phonon dispersion and phonon density of states of HSH-carbon. Average potential energy fluctuation of HSH-carbon during ab initio NpT ensemble simulations under the temperature of **(c)** 500 K and **(d)** 800 K with top and side views of structures at the end of the simulations as the inset.

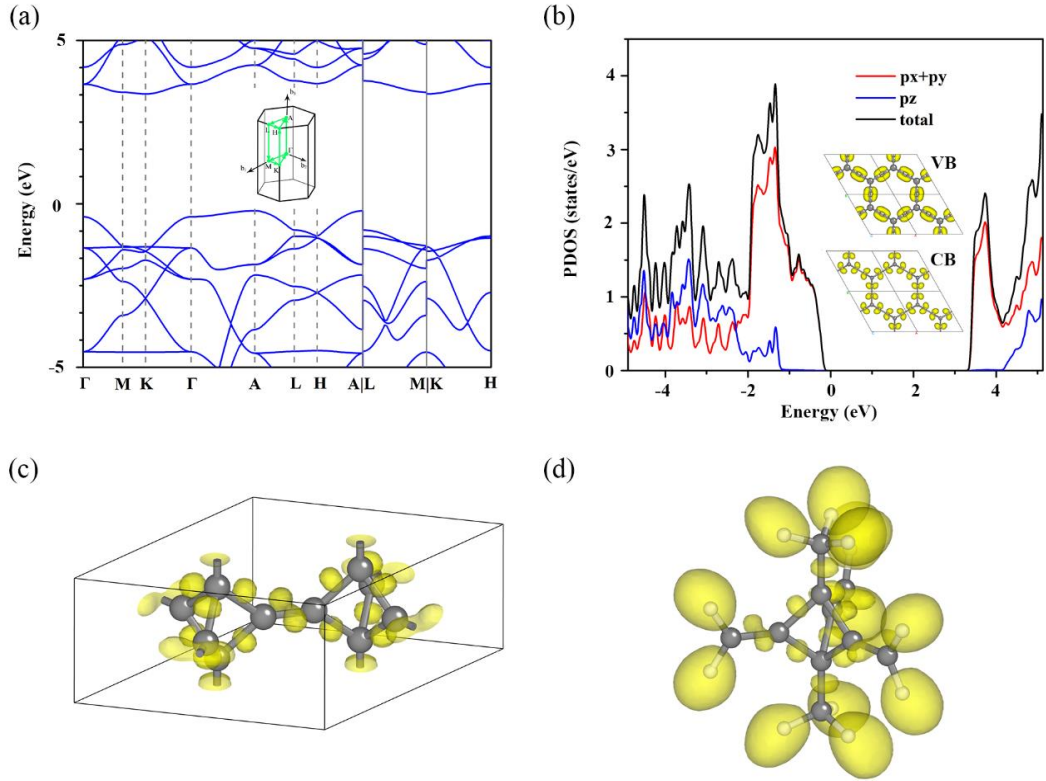

**Fig. 3 (a)** The electronic band structure and its corresponding Brillouin zone path. **(b)** Projected density of states (PDOS) with Fermi level shifted to 0 eV, with local density of states for the highest branch of valence bands (VB, isosurface value of 0.08 e/Å$^3$) and the lowest branch of conduction bands (CB, isosurface value of 0.18 e/Å$^3$) at Γ point as the inset. **(c)** The electron localization function (ELF) of HSH-carbon (isosurface value of 0.863). **(d)** The ELF of a partial methylated [1.1.1]propellane molecule (isosurface value of 0.863). The solid lines in the structure represent the periodical boundary. The isosurfaces of VB, CB and ELF are obtained with Device Studio [33].

**Table 2.** Effective mass of carriers-electrons and holes of HSH-carbon.

|  | Effective mass along K-path (in units of free electron mass m$_0$) | | | |
| --- | --- | --- | --- | --- |
| CB edge | K→M | K→Γ | Γ→M | Γ→K |
|  | 2.90 | 2.41 | 0.481, -2.34 | 0.485, -2.37 |
| VB edge | A→L | A→H | Γ→M | Γ→K |
|  | -1.60 | -1.60 | -0.898 | -0.903 |

**Table 3.** Elastic properties, including bulk modulus ($B$), Young's modulus ($E$), shear modulus ($G$) and Poisson's ratio ($v$) of c-diamond, T-carbon and HSH-carbon based on Voigt-Reuss-Hill (VRH) approximation.

| Material | Reference | $B$ (GPa) | $E$ (GPa) | $G$ (GPa) | $v$ |
|---|---|---|---|---|---|
| c-diamond | Ref. [11] | 464 | 1100 | 522 | 0.070 |
| T-carbon | Ref. [11] | 169 | 185 | 70 | 0.318 |
| HSH-carbon | this work | 170 | 120 | 43 | 0.384 |

**Table 4.** Elastic properties and anisotropy of HSH-carbon.

| Elastic properties | | HSH-carbon |
|---|---|---|
| Young's modulus, $E$ (GPa) | $E_{max}$ | 530.76 |
| | $E_{min}$ | 69.03 |
| | $A_E = E_{max} / E_{min}$ | 7.69 |
| Shear modulus, $G$ (GPa) | $G_{max}$ | 60.42 |
| | $G_{min}$ | 18.64 |
| | $A_G = G_{max} / G_{min}$ | 3.24 |
| Linear compressibility, $\beta$ | $\beta_{max}$ | 2.07 |
| | $\beta_{min}$ | 1.70 |
| Poisson's ratio, $v$ | $v_{max}$ | 0.85 |
| | $v_{min}$ | 0.01 |
| | $A_v = v_{max} / v_{min}$ | 85.00 |

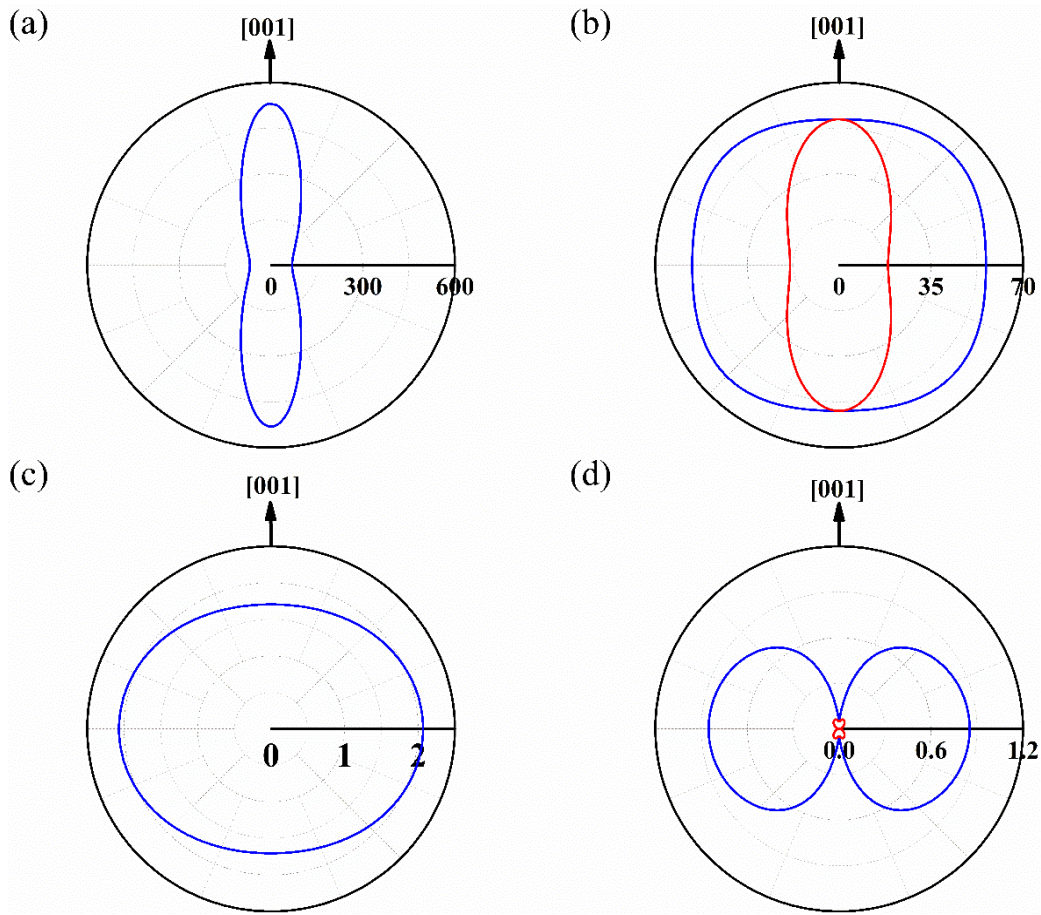

**Fig. 4** Projection of 3D representation in a plane along the <001> orientation of **(a)** Young's modulus, **(b)** shear modulus, **(c)** linear compressibility and **(d)** Poisson's ratio of HSH-carbon. For figures with multiple curves, blue and red represent the maximum, minimum of the quantities, respectively.